\documentclass[pdftex, twoside, dvipsnames]{article}
\usepackage{amsmath,amssymb,amsthm,amsfonts}
\usepackage{bm}  % From amsmath, a bold font

\newtheorem{thm}{Theorem}

\everymath{\displaystyle}
\allowdisplaybreaks
\usepackage{mathtools}
\usepackage[makeroom]{cancel}
\usepackage{newtxtext,newtxmath}
\usepackage[title]{appendix}
\usepackage{epigraph}
\usepackage{fancyhdr}
\fancypagestyle{firstpagestyle}{

\fancyhf{} % clear all header and footer fields
\fancyfoot[C]{\thepage}
}
\fancypagestyle{CVfirstpagestyle}{

\fancyhf{} % clear all header and footer fields
\fancyfoot[C]{\thepage}
}
\usepackage{natbib}
\newcommand{\Rea}{\mathbb{R}}

\newcommand{\Pe}{\mathbb{P}}

\title{Climate uncertainty, financial frictions and constrained efficient carbon taxation\thanks{I thank William Brock, Harold Cole and Herakles Polemarchakis for very useful conversations and comments. This work was supported by the Swiss National Science Foundation (SNF), under project ID  \lq\lq Can Economic Policy Mitigate Climate-Change?\rq\rq.}}

\author{
  Felix K\"ubler\thanks{Department for Banking and Finance, University of Z\"urich; Swiss Finance Institute (SFI); Email: fkubler@gmail.com}}
\date{\today}

\begin{document}
\maketitle

\begin{abstract}
 In this paper, I consider a simple heterogeneous agents model of a production economy with uncertain climate change and examine constrained efficient carbon taxation. If there are frictionless, complete  financial markets, the simple model predicts a unique Pareto-optimal level of carbon taxes and abatement. In the presence of financial frictions, however, the optimal level of abatement cannot be defined without taking a stand on how abatement costs are distributed among individuals. 
 I propose a simple linear cost-sharing scheme  that has several desirable normative properties.
 I use calibrated examples  of economies with incomplete financial markets and/or limited market participation to 
 demonstrate that different schemes to share abatement costs can have large effects on optimal abatement levels and that
 the presence of financial frictions can increase optimal abatement by a factor of three relative to the case of frictionless financial market.
\end{abstract}
\clearpage
\section{Introduction}

The negative effects of anthropogenic climate change will differ dramatically across regions and across different individuals within regions (see, e.g., \cite{nordhaus1996regional}, \cite{dennig2015inequality}, \cite{Cruz2021}, \cite{KKPS}, \cite{krusell2022climate}). Moreover, the extent of future damages is highly uncertain (see, e.g., \cite{hassler2018consequences} or \cite{barnett2022aclimate}). Many agents do not have access to financial markets, and existing markets are incomplete in that climate uncertainty cannot be hedged. These financial frictions, together with agents' heterogeneity, obviously exacerbate welfare losses from climate change.
Carbon pricing can reduce emissions and the level as well as the uncertainty about future climate-related damages. 
However, there is obviously a trade-off between the costs of abatement and the future costs of climate change. Normative statements about the optimal level of abatement are only possible if one takes a stand on how the abatement costs should be shared across agents.
Financial frictions can prevent those agents who suffer the most from future damages from using resources from those future states with little damage to combat climate change today. Depending on the abatement cost-sharing scheme, carbon policy might only help little to combat welfare losses from climate change.

In this paper, 
I consider a simple stochastic growth model with heterogeneous agents and climate change to formalize these ideas. Financial frictions take the form of incomplete financial markets and restricted participation in the existing markets.
The primary goal is to find a constrained Pareto-efficient path of carbon-risk taxes and abatement cost sharing  and to show how they differ from the ones in an economy without financial frictions. I also examine the welfare losses that are caused by a combination of financial frictions and climate damages.

This is hardly the first quantitative analysis of uncertainty's importance to optimal carbon policy. Prior major studies include ~\cite{jensen2014optimal},~\cite{cai2013social}, and~\cite{traeger2018ace} who all focus on uncertainty arising from tipping points. More closely related to this paper's treatment of uncertainty are \cite{hassler2018consequences}, \cite{van2021risk}, \cite{kotlikoff2020pareto}, \cite{barnett2020pricing}, \cite{barnett2022aclimate} and \cite{barnett2022climate}\footnote{See, e.g.,~\cite{gillingham2015modeling} and ~\cite{cai2020role} for thorough reviews of the literature.}. 
This said,  this paper appears to be the first study of constrained optimal carbon-risk tax policy in a heterogeneous agents model with financial frictions. 
Given the importance of the abatement cost-sharing scheme, this paper is also closely related to the literature on optimal carbon taxation and redistribution in economies with heterogeneous agents (see, e.g., \cite{barrage2020optimal} and \cite{douenne2022optimal}). This literature focuses largely on who bears the burden of carbon taxes, while in my paper, the problem is  modeled in an overly simplified fashion.
There is also literature on the role of financial frictions in the amplification of transition risk (see, e.g., \cite{carattini2021climate}). While financial frictions also play a role in my paper, there is no close connection.

In the model, higher temperature negatively affects labor productivity. These effects differ in magnitude across different agents. 
 Future temperatures are uncertain because of uncertainty about the climate sensitivity parameter. Markets are incomplete in that agents can only trade risky capital. In addition, some agents might have no access to asset markets whatsoever.
 
The supply side of the model is a slightly simplified version of the model in \cite{nordhaus2017revisiting}. Final goods are produced using capital, labor, clean energy, and dirty energy, emitting CO2 into the atmosphere. Importantly, there is a single capital stock. This makes an interpretation of the heterogenous agents in the economy as agents in different regions somewhat problematic. In a model with region-specific production functions, one would need to take a stand on capital mobility and international financial markets, which is a bit beyond the scope of this study.

Abatement can reduce CO2 emissions at some reduced form cost function. 
    I adopt the carbon cycle and temperature equation from~\cite{golosov2014optimal}, which simplifies~\cite{nordhaus2017revisiting}'s treatment of these elements.\footnote{ 
As \cite{Folini_2021} point out, this model misses various important features of climate change, but for the purposes of this study,  it seems an acceptable simplification. In particular, the length of a period is taken to be 10 years which alleviates some of the issues raised in \cite{Folini_2021}. }

The uncertainty about the climate-sensitivity parameter (ECS) is the key ingredient of the model.   While the climate-sensitivity coefficient plays a crucial role in understanding the effects of CO2 emissions on global warming, there is, unfortunately, little scientific consensus on its value. As~\cite{knutti2017beyond} put it, \lq\lq \textit{Equilibrium climate sensitivity characterizes the Earth's long-term global temperature response to increased atmospheric CO2
concentration. The temperature response has reached almost iconic status as the single number that describes how severe climate change will be. Interestingly, the consensus on the ‘likely’ range of temperature increase -- 1.5  to 4.5 degrees Celsius -- is the same today as the projection made by Jule Charney in 1979.}\rq\rq 
 A large part of the uncertainty about the ECS is typically attributed to model uncertainty. In other words, researchers believe that the time variation in the true parameter is much smaller than this range but that our current climate models cannot determine this true parameter. However, there is some evidence that even beyond model uncertainty, the equilibrium climate sensitivity might vary over time (see, e.g.,~\cite{roe2007climate} but also \cite{Zaliapin.Ghil.2011}).  Based on my reading of the climate-science  literature, I model this parameter as a Markov process whose persistence increases over time so that in 300 years, the model is essentially deterministic.

The policy instruments are i) a tax on dirty energy's use that depends on time and the state of the economy and ii) a sharing rule (also varying with time and the state) of each period's abatement costs. Critically, I do not allow for arbitrary transfers but only for those that compensate agents for their losses caused by abatement.
Because of financial frictions, there is no Pareto-efficient policy using only these two instruments. 
The focus of this paper is  on \lq\lq constrained interim efficient\rq\rq \ policies. In each period, current carbon-taxes are Pareto-efficient  in the sense that there are no other taxes and transfers in the period that make everybody better off.
On one hand, ex-interim efficiency seems to be a reasonable requirement since each period, taxes and transfers are optimal in that period.
On the other hand, in the presence of financial frictions, one can increase everybody's ex-ante welfare by dropping the interim efficiency requirement, even if transfers are restricted to only compensate for abatement costs. If agents pay more in states where they are relatively rich and less in states where there are relatively poor, since marginal rates of substitution differ across agents, everybody's ex-ante welfare can increase. In this paper, possible ex-ante Pareto-improvements are ignored, and the focus lies on ex-interim efficiency.
The concept is referred to as interim efficiency since it can be interpreted as a generalization of ex-post efficiency to a multi-period model. \cite{starr1973optimal} already notes the tension between ex-ante and ex-post efficiency in a different context.
 This problem is also somewhat similar to issues arising in time-consistent public policy models (e.g., \cite{klein2008time}). 
I do not explicitly model a policy game, but one reason for the constraints on policies to be ex-interim efficient can lie in a lack of commitment.

Of course, there are infinitely many policies that satisfy the condition of constrained interim Pareto-efficiency, and they all differ with respect to abatement levels and transfers.
I focus on a particular one, where each period, all agents are compensated exactly for the losses due to abatement, and the transfers necessary for this are shared proportionally to agents' marginal willingness to pay for the reduction in CO2 emissions. In other words, prices are held fixed at their pre-abatement levels, and abatement costs are shared according to a linear rule. The concept can be viewed as a variation of a  {\it Balanced Linear Cost Share Equilibrium} as in \cite{mas1989cost} or a {\it Ratio Equilibrium} in the sense of \cite{kaneko1977ratio}.
As \cite{mas1989cost}
point out,  \lq\lq no justification in terms of fairness should be attempted for this distribution scheme\rq\rq \ (see also \cite{moulin1987egalitarian} for cost-sharing schemes that can be justified by normative principles), but there are a variety of practical arguments in its favor.
It is easy to prove that this policy satisfies the definition of constrained efficiency and that, given the transfer rule, there are no other taxes that lead to an ex-ante Pareto-improvement.
By construction, the policy constitutes a Pareto-improvement relative to the laissez-faire outcome. Most importantly, it is tractable in that optimal abatement and transfers can be calculated relatively easily.
In a slight abuse of language, this policy is sometimes referred to as {\it the constrained optimal}, or constrained efficient policy.

I compare the policy to a policy where taxes are set to be interim efficient, but there are no transfers. It is easy to see that in this setup, these taxes will generally not be ex-ante efficient. The resulting allocation might not constitute a Pareto-improvement over the laissez-faire outcome.
These taxes are generally higher than the constrained optimal taxes, and the hand-to-mouth consumers that are hurt most by climate change are better off, while the consumers that are hurt least by climate change are worse off.
However, the effects on welfare are quantitatively small in our calibration.

I also compare the constrained efficient policy to the optimal policy in a world without financial frictions. Optimal taxes are significantly lower by a factor of 2-3, (and as a result, future temperatures and damages are higher) in the absence of financial frictions. This might seem counter-intuitive at first, but is caused by the precautionary savings motive resulting from uncertainty.

In the calibrations, the welfare properties of the constrained efficient policy depend crucially on abatement costs. For low abatement costs, the welfare losses due to financial frictions are very small, about two percent of certainty equivalent lifetime consumption for  the agents hit most by climate change. If abatement costs become very high, overall welfare losses obviously increase but, more importantly, welfare losses from financial frictions increase as much as welfare costs from emissions.

The remainder of the paper is organized as follows: Section~\ref{sec:model} presents the basic model.
In Section ~\ref{sec:ceff} I introduce constrained efficient carbon policy
Section~\ref{sec:cali} discusses the calibration strategy and Section \ref{sec:results} presents the results.
Section \ref{sec:concl} concludes. An appendix discusses some computational details.

\section{Model}
\label{sec:model}
In this section, the economic model is introduced. The climate change externality is described and economic policy is defined.

\subsection{The physical economy}
This paper considers a production economy with long-lived agents. 
Time is indexed by $t =1,\ldots,T \le \infty $. Exogenous shocks $ z_t $ realize in a finite set $ {\mathbf Z}=\{1,\ldots,Z\} $, and follow a first-order Markov process with transition probability $ \Pe_t(.|z) $ at each date $ t $ .
 A history of shocks up to some date $ t $ is denoted by $ z^t = (z_0,z_1,\ldots,z_t)$ and called a date event. Whenever convenient, I simply use  $ t $ instead of $z^t$. 

At each date-event there is a single good available for consumption. Capital can be stored and agents are endowed with labor. 
%%%%%%%%%%%%%%%%%%%%%%%%%%%%%%%%%%%%%%%%%%%%%%%%%%%%%%%%%%%%%%%%%%%%%%%%%%%%%%%%%%%%%%
\subsection{Firms}
\label{sec:firms}
%%%%%%%%%%%%%%%%%%%%%%%%%%%%%%%%%%%%%%%%%%%%%%%%%%%%%%%%%%%%%%%%%%%%%%%%%%%%%%%%%%%%%%
 A representative firm produces the consumption good using capital and labor.
The production function is Cobb-Douglas and given by
\begin{equation} 
\label{eq1} 
f(K_t,l_t,\mu_t) = A_t(\mu_t)  K_{t}^{ \alpha} L_{t}^{(1-\alpha)}
+(1-\delta)K_{t}, \end{equation}
where   $A_{t}$,  $K_{t}$, $L_{t}$ refer to total factor productivity,  capital and labor respectively. The parameter $\alpha \in (0,1)$ represents the capital share, $ \mu_t \in [0,1] $ denotes abatement and $ \delta $ the depreciation rate. Throughout it is abstracted from TFP growth and I take $ A_t(0) = 1 $ for all $ t$. It is useful to define $ Y_t=K_{t}^{ \alpha} L_{t}^{(1-\alpha)} $ as potential output without any abatement.
I assume throughout that for all $t$, $ A_t $ is decreasing, concave and differentiable and that $ A_t(0)=1 $.

Production emits CO2 into the atmosphere. 
I take as exogenous a sequence  $ (e_t)_{t=0}^T  $ and assume that total emissions in period $t$ are given by
$ E_t = (1-\mu_t) e_t Y_t $, where $ \mu_t $ is the level of abatement that reduces production via $ A_t(\mu) $. 

Firms rent capital and buy labor on spot markets, facing a wage $w_t$ and a price of capital $ r_t $. Since they make no intertemporal decisions, there is no issue about the appropriate objective function in a model with financial frictions.
This is an important simplification that is crucial for the analysis; in a more realistic setup that models the use of oil, gas, and coal  explicitly, this is unlikely to be the case, and without taking a stand on the theory of the firm in incomplete markets, one cannot talk about optimal policy (see, e.g., \cite{dierker2002nonexistence}, for some of the complications that arise). 

%%%%%%%%%%%%%%%%%%%%%%%%%%%%%%%%%%%%%%%%%%%%%%%%%%%%%%%%%%%%%%%%%%%%%%%5     
\subsection{Households}
\label{sec:household}
%%%%%%%%%%%%%%%%%%%%%%%%%%%%%%%%%%%%%%%%%%%%%%%%%%%%%%%%%%%%%%%%%%%%%%%5     
There are $H$ types of $ T$-period lived agents, $h \in {\mathbf H}=\{ 1,\ldots,H \}$,
 that maximize lifetime expected utility, given by
\begin{equation}
U(C) = {\mathbb E}_0 \sum_{t = 0}^T \beta^t \frac{c_{t}^{1-\sigma}-1}{1-\sigma},
\end{equation} 
subject to
\begin{equation}
c_{t}+a_{t+1} =(1+r_t)a_{t}  +w_t l_{h,t} -\theta_{t,h},\ a_t \in {\mathbf Y}^h
\label{eq:BC}
\end{equation} 
where $\beta \in (0,1) $ is the time preference factor, $ \sigma>0 $ denotes the coefficient of relative risk aversion, $c_{t}$, $a_{t}$, $ r_t$, $w_{t}$ correspond to consumption, capital holdings, returns to capital and wages  at time $t$, respectively, and labor supply at $t$ is $l_{h,t}$. Capital holdings of agent $h$ are restricted to lie in a set $ {\mathbf Y}^h $ -- some agents might not be able to participate in capital markets at all.
In the analysis below, I also consider the case where,
in addition to risky capital, there is a complete set of Arrow securities available for trade and where no agents face restrictions on their trades. I refer to this case as complete markets and contrast it to the economy with financial frictions described here.

The variable $ \theta_{t,h} $ denotes the possibly state-specific net tax paid by the agent $h$ at time $t$. In our specification of transfers below this can depend on the agents' beginning of period capital holdings, a fact the agent takes into account when making savings decisions.
It is assumed throughout that $ \sum_h \theta_{t,h}=0 $ for all $t$.

I assume that an agent's labor productivity is impacted negatively  by climate change. For each agent $ h \in {\mathbf H} $ there is a agent-specific damage function $D_{h}(T_t)$, where $T_t$ denotes average surface temperature (in degrees Kelvin) relative to pre-industrial levels.
An agent's labor supply at time $t$ is denoted by
$ l_{t,h}=D_h(T_t) \bar{l}_h $, where $ \bar{l}_h>0 $ can be thought of as the agents labor supply without climate change.
Aggregate damages to labor supply are denoted by
$$ D(T_t) = \sum_{h \in {\mathbf H}} D_h(T_t) \bar{l}_h ,$$
and given the assumption on Cobb-Douglas production, they translate to an effect of total output that can be written as
$ D(T_t)^{1-\alpha} $.

%%%%%%%%%%%%%%%%%%%%%%%%%%%%%%%%%%%%%%%%%%%%%%%%%%%%%%%%%%%%%%%%%%%%%%%5     
\subsection{Modeling climate change}
%%%%%%%%%%%%%%%%%%%%%%%%%%%%%%%%%%%%%%%%%%%%%%%%%%%%%%%%%%%%%%%%%%%%%%%5     

 The carbon cycle is modeled as in~\cite{golosov2014optimal}. The temperature $T_t$ in period $t$ is determined by the stock of carbon in the atmosphere, $ S_t $,
\begin{equation} 
\label{forcing}  T_t=\lambda_t \frac{\log(S_t/S)}{\log(2)} ,
\end{equation}
where $ S $ is the pre-industrial carbon stock. We model $ \lambda_t $ in a stochastic manner, assuming that $ \lambda_t $ only depends on the current shock $ z_t $.

Following~\cite{golosov2014optimal}, it is assumed that the CO2 stock in the atmosphere has two components---that is, 
\begin{equation}
S_t=S_{1t}+S_{2t} ,
\label{eq:S1}
\end{equation}
where
\begin{equation}
S_{1t}=\xi_1  \xi_2 \cdot  E_t+ \delta_{S1} \cdot S_{1,t-1},
\label{eq:S2}
\end{equation}
and where 
 \begin{equation}
S_{2t}=\xi_1 (1-\xi_2)\cdot \xi \cdot  E_t + \delta_{S2} \cdot  S_{2,t-1}.
\label{eq:S3}
\end{equation}
The depreciation parameters satisfy $ \delta_{S2} < \delta_{S1} \le 1 $. I calibrate the former at a low value and the latter at a high value, again following~\cite{golosov2014optimal}. Hence, $S_{1t}$ is a slowly depreciating stock of carbon, whereas $S_{2t}$ is a rapidly depreciating stock. The parameters $ \xi_1 $ and  $ \xi_2 $ control the fraction of CO2 emissions entering the atmosphere. We take $ \xi_1, \xi_2 $, and the depreciation parameters as fixed.  As~\cite{golosov2014optimal} point out, there is no consensus in the literature concerning the values of these parameters. \cite{Folini_2021} discuss better models for the carbon cycle in detail.
In any case, the main results in this paper are robust to moderate differences in these parameters as well as to time-varying shocks they may experience. 
    
Climate change reduces output productivity by destroying agents' labor endowments, as described above.
    
\subsection{Abatement}
Firms can reduce CO2 emissions via abatement, and a carbon tax will lead to a reduction in emissions that depends on the abatement technology. Instead of modeling dirty energy supply explicitly, I use the reduced form formulation from \cite{nordhaus2017revisiting}.
As mentioned above, it is assumed that emissions $ E_t $
are given by the level of production, an exogenous time-varying parameter, and abatement $ \mu $, and therefore the abatement necessary to achieve emissions $ E_t $ in period $t$ can be written as 
$$ \mu_t = 1-\frac{E_t}{e_t Y_t} . $$
Given a tax $ \tau $ per unit of CO2 emissions firms choose $ \mu $
so that profits are maximized. The first order condition for optimal abatement is  therefore given by
$$ A_t'(\mu_t)\frac{1}{e_t Y_t}  Y_t = \tau_t ,$$
which  give a unique $ \mu_t(\tau) $ for each $ \tau $.
Instead of searching for a tax policy, one can, therefore, directly search for optimal abatement, $ \mu_t $. The revenue from the carbon tax, $ \tau E_t $ is assumed to be redistributed lump-sum to the firms so that transfers only have to compensate for the abatement costs,  $(1-A_t(\mu_t)) Y_t $.

\subsection{Policy}
A climate policy taxes emissions and institutes transfers between agents to compensate them for the losses incurred by taxation.
As explained above, I will formally write this policy as abatement levels and transfers but will often refer to it as taxes and transfers.

In the setup of this paper, it is natural to assume that taxes can depend on time and the state of the economy (i.e., the exogenous shock as well as CO2 concentrations and endogenous economic variables). Since the period length amounts to 10 years, it would be unreasonable to assume that with the arrival of new information regarding the ECS and CO2 concentrations, taxes and transfers cannot be adjusted.
Agents take as given a tax- and transfer function that depends on time, the shock, and the state of the economy.
The endogenous state, $ s $, consists of beginning-of-period capital holdings across agents, $(a_1,\ldots,a_H)$ and beginning-of-period CO2 stocks $ (S_1,S_2) $.
I write $ \mu_t(z,s)  $ and $ (\theta_{h,t}(z,s,\mu))_{h \in {\mathbf H}} $ to denote the policy. Crucially, taxes are assumed not to depend on individual variables, i.e. a single household's choices at $ t-1 $ do not affect taxes.
Transfers, on the other hand, will compensate for lower returns due to abatement and, therefore, will depend on agents' choices at $ t-1 $. As mentioned above, I directly work with abatement $ \mu $ instead of taxes $ \tau $ since this simplifies the notation. In a slight abuse of language,  $ \mu $ is sometimes referred to as the tax.
\subsection{Equilibrium}
Given taxes and transfers, a competitive equilibrium consists of a stochastic process for choices and prices  so that asset markets and labor markets clear and all agents  maximize utility given the tax and transfer policies. In this specification this implies that $ K_t= \sum_h a_t^h $, $ L_t= \sum_h l_{ht} $ and wages and the rental rate of capital are given by the marginal product of labor and capital in the firm's production function (given optimal abatement). I sometimes write $ r(\mu) $ and $ w(\mu) $ to make explicit the dependence on abatement.

Throughout the analysis, it is assumed that taxes and transfers are a function of time, the current shock, and the endogenous state and that 
there exists a recursive equilibrium all optimal choices  are functions of the current shock, the endogenous state, and time, $t$, alone.
The endogenous state, $ s $, consists of beginning-of-period capital holdings across agents, $(a_1,\ldots,a_H)$ and beginning-of-period CO2 stocks $ (S_1,S_2) $.
An agent's value function, $ V^h_t(z,s,a) $ gives his equilibrium utility as a function of time, the aggregate state, i.e, the endogenous state, and the current shock, $ z_t $, and the agent's beginning of period asset holding, $a$.
It is useful to define the Q-function as the value also depending on current policy choices, i.e. for every agent $h$,
$$ Q^h(z,s,a; \mu, \theta) = \max_{a'} u(c) + \beta E_t V^h(z',s',a') \mbox{ s.t. } $$
$$ c=a(1+r(\mu)) + l_{h}(T(s)) w(\mu) -a' - \theta_{h}  $$
$$ S_1'=\xi_2 \cdot \xi_1 \cdot  (1-\mu) Y_t e_t + \delta_{S1} \cdot S_{1},\ 
S_{2}'=(1-\xi_2)\cdot \xi_1 \cdot (1-\mu)  Y_t e_t  + \delta_{S2} \cdot  S_{2}.$$

It is assumed throughout that a recursive equilibrium exists -- even in simple models without climate change 
this might be a strong assumption (see, e.g. \cite{brumm2017recursive}).
However, clearly, one needs to avoid the possibility of multiple equilibria in the comparative statics exercises.

\section{Constrained efficient policy}
\label{sec:ceff}
In standard economic models, climate change is treated as an externality, and, going back all the way to the work of~\cite{pigou1920wealth}, it is well known that Pareto efficiency can be restored by adding the marginal market value of the externality to the market price of carbon.
In settings where the welfare theorems hold, this can be shown in some generality, although there is, of course, an issue with uniqueness.
In this setting, this value cannot be deduced from market prices and has to be obtained by adding up each agent's marginal valuation. Since the welfare theorems fail due to financial frictions,  a notion of constrained efficiency has to be employed in order to examine carbon taxes that have desirable normative properties.

As explained above, taxes are functions of time and the exogenous and endogenous state of the economy.
In the first best,  transfers would also depend on the state and lead to a Pareto-optimal allocation. However, since it is assumed that there are no financial assets that pay contingent on the realization of the climate shock, it seems reasonable to assume that  transfers cannot compensate for the lack of asset markets.  I have in mind a model where at time $t$, the value of $ \lambda_t $ is observable but where it is not ex-ante (i.e. at any time before $t$) contractible. At time $t$, once $ \lambda_t $ is observed taxes can be adjusted accordingly, but transfers are restricted to compensate only for losses that are caused by taxes in the current period.
Pareto efficiency can generally not be obtained with such a scheme, and, in general, it is difficult to characterize constrained efficient taxes and transfers.
Since markets are incomplete, the tax policy will have an insurance aspect that depends on taxes across states in the future.
The key assumption in this paper is that policy is restricted to be ex-interim efficient.
Formally, we say that
a tax- and transfer policy is interim constrained Pareto-efficient  if at any node $ z^t $,
there are no taxes and transfer $ (\widetilde \mu, \widetilde \theta) $ such that for all $ h \in {\mathbf H} $
$$ Q^h_t(z_t,s(z^t),\widetilde{\mu} ,\widetilde{\theta}) \ge
V^h_t(z_t, s(z^t)) , $$
where the inequality holds strict for at least one agent.

The requirement of interim efficiency puts obvious restrictions on taxes.
An interim-efficient tax solves
$$ \max \sum_h \lambda_h Q^h(z,s,\mu,\theta) 
\mbox{ s.t. } \sum_h \theta_h=0 $$
for some weights $ \lambda \in \Rea^H_{++} $.
 It is easy to see that
 a necessary condition for a constrained interim efficient $ \mu $ is as follows.
\begin{equation} \label{optax}
  \sum_{h\in {\mathbf H}} \frac{\frac{\partial Q^h}{\partial \mu}}{u'(c^h_t)} = 0. \end{equation}
 The sum of the marginal benefits of an infinitesimal increase in abatement must equal the marginal cost.
In the general equilibrium setting of this paper, the condition will generally not be sufficient to determine the optimal level of abatement uniquely.
In the following, I make the strong assumption that there exists a unique solution to this equation that is denoted by  $ \mu^*_t(z,s) $.

While constrained optimal taxes must satisfy (\ref{optax}), they obviously depend crucially on the welfare weights $ \lambda\in \Rea^H_+ $, which have been substituted out of the expression and are replaced by a transfer scheme. Without specifying the welfare weights,
there are obviously infinitely many possible transfers, even if one focuses on a system of transfers that is interim Pareto-improving relative to the laissez-faire outcome. Moreover, the levels of constrained optimal taxes and of abatement depend critically on the transfer scheme, and it is impossible to characterize optimal taxes without, at the same time, characterizing transfers.

In general, this is to be expected in a model with heterogeneous agents -- the distribution of wealth affects asset prices and hence the evaluation of future damages.
However, in this paper, it is assumed that agents have identical CRRA utility functions, and in this setup, optimal abatement is independent of transfers when markets are complete (see, e.g. \cite{gollier2001wealth}).
With financial frictions (incomplete financial markets and/or limited participation) optimal abatement always depends on transfers. The agents' first-order conditions with respect to asset holdings are (generically) linearly independent of Equation (\ref{optax}) and the solution of $ (\ref{optax}) $ depends on the distribution of income in the current period.
A calibrated example in Section \ref{sec:results} below demonstrates that different cost-sharing rules can imply significant differences in optimal abatement.

The joint determination of optimal abatement and its cost-sharing is formally equivalent to a model of the optimal supply of a public good (at least in the simple model considered in this paper, see \cite{chichilnisky1994should} for a discussion of the problem in a more sophisticated setup with heterogeneous produces and heterogeneous consumers).
The classical formalization of a Lindahl equilibrium in
\cite{foley1970lindahl}  takes ownership shares (in the abatement technology) as given and is therefore not suitable for our setting. \cite{kaneko1977ratio} and \cite{mas1989cost} do not consider ownership shares and
derive a linear cost-sharing rule which I adopt in this study. The scheme fits well with the idea that individual contributions to the cost reflect personal valuations and is very tractable compared to other approaches (e.g. \cite{moulin1987egalitarian}).

To formulate the scheme, define future benefits from abatement today as
$$ \tilde{Q}^h_t(\mu) = \frac{\partial Q_t^h}
{\partial \mu} - u'(c^h_t) A_t'(\mu) \left(l^h_t (1-\alpha)K_t^{\alpha} L_t^{-\alpha}
+k^h_t \alpha K_t^{\alpha-1} L_t^{1-\alpha} \right) 
$$
and set transfers to be 
\begin{eqnarray} \label{optran}  \theta_{th}(z,s) &=& \frac{\tilde{Q}^h_t(\mu^*_t(z,s))}
{\sum_i\tilde{Q}^h_t(\mu^*_t(z,s))}
(A(\mu^*_t(z,s))-A(0)) K_t^{\alpha} L_t^{1-\alpha}
-\\
\notag
&&
(A(\mu_t)-A(0)) \left(l^h_t (1-\alpha)K_t^{\alpha} L_t^{-\alpha}
+k^h_t \alpha K_t^{\alpha-1} L_t^{1-\alpha} \right) .  \end{eqnarray}

In the scheme,
capital holders are compensated for the lower
returns, and workers are compensated for the lower wages caused by abatement. The capital holders take this into account when making investment decisions at $ t-1 $. The costs of abatement are shared according who gains most from the abatement irrespective of the initial endowments in capital and labor.
In a model with technical change, where agents have 
an option to invest in dirty versus clean energy technologies, this compensation causes an obvious inefficiency. In this paper, this is not the case, and, as mentioned above, a more realistic supply side would entail taking a stand on the problem of the policy of the firm with financial frictions. Something that needs to be avoided.
It seems that it is important to leave asset returns unchanged through the transfers. Changes in prices lead to changes in savings, and in the incomplete markets setup considered here, that alone generally has welfare consequences (see \cite{davila2012constrained}). As mentioned above, a collection of taxes and cost-sharing across nodes that is ex-ante efficient would take all of this into account and might result in abatement levels that are very different than the ones reported below. However, it is important to emphasize that the concept here is purely forward-looking.

As it is shown in Theorem \ref{thm1} below, this scheme of taxes and transfers has several desirable properties.
It is an interim improvement relative to the laissez-faire, that is, at all $ z^t$, for all $ h \in {\mathbf H} $, $$ V^h(z_t , s(z^t)) \ge Q^h_t(z_t,s(z^t),0,0) .$$
Moreover, given transfer functions $ \theta(s,z,\mu) $, the tax policy is ex ante efficient, i.e. 
 there is no other policy, $ \tilde{\mu}$ that satisfies 
for all $ h \in {\mathbf H} $
$$ Q^h_0(z_0,s_0,|(\widetilde{\mu})) \ge
V^h_0(z_0, s_0) , $$
where the inequality is strict for at least one agent.

The following theorem summarizes the main results of this section.
\begin{thm}
\label{thm1}
Suppose there exists a recursive equilibrium and suppose Equation (\ref{optax}) always has a unique solution. Then 
\begin{itemize}
\item The policy given by  (\ref{optax}) and (\ref{optran}) is  constrained interim efficient.
\item It constitutes an interim Pareto-improvement relative to the laissez-faire equilibrium.
\item 
There is no other tax policy $ (\tau(z^t))_{z^t} $ with transfer given by  (\ref{optran}) that makes all agents ex ante better off.
\end{itemize}

\end{thm}
The first two properties follow directly by the construction of the policy. Ex-ante efficiency is guaranteed because each agent's welfare at each $t$ reaches its maximum at $ \mu^* $. Any change in taxes must make at least one agent worse off.
By construction, the linear cost-sharing scheme implies that at each $t$ all agents agree on the optimal level of abatement.

As mentioned in the introduction, the policy defined by (\ref{optax}) and (\ref{optran}) is sometimes referred to as the {\it constrained optimal policy} in this paper.

The assumption that the $Q$ function is convex in emissions is implicit in the assumption that there is a unique solution to Equation (\ref{optax}) . It is the analog to a standard convex cost assumption. In the case of climate change, this assumption is more problematic, however, even in a representative agent model. There are two competing supply-side non-linearities in climate models. First, damages are assumed to be a strictly convex function of global average surface temperature. Second, the average surface temperature is modeled as a logarithmic and, therefore, a concave function of atmospheric CO2. Often the convexity outweighs its supply-side concavity in determining carbon-risk damages. Moreover, when it comes to the welfare effects, risk aversion is likely to outweigh the concavity in the forcing equation.

I compare this transfer scheme with constrained efficient taxation without transfers below. Note that without transfers, taxes will be ex-interim constrained efficient if they satisfy \ref{optax}. In the presence of financial frictions, they will generally not be ex-ante efficient, and they might not lead to an allocation that Pareto dominates the laissez-faire.

I also compare the {\it constrained optimal} policy to the optimal policy in an economy without financial frictions. In this case, the optimal tax is simply set to the discounted (at the stochastic discount factor) sum of future damages caused by one extra infinitesimal unit of emissions in the current period. Since agents are assumed 
to have identical CRRA utility, this so-called social cost of carbon does not depend on who bears the abatement cost. To determine optimal taxes, one can simply use a representative agent economy. Transfers can then be chosen independently to trace out the Pareto-frontier of the economy.

\section{Calibration}
\label{sec:cali}
I take a model period to be 10 years and assume $ T=30 $ -- a horizon of 300 years is convenient for calibrating emissions, and a model period of 10 years is the same as in \cite{golosov2014optimal}. A longer, even infinite time horizon has only small effects on optimal abatement and welfare in the first periods.

Technology and preferences are standard. It is assumed that the capital share in the aggregate production function is $ \alpha=0.33 $. The capital depreciation rate, $  \delta$ is set to 0.57 (corresponding to 8 percent depreciation annually). 
For simplicity, there is no TFP growth, and absent climate change the economy would be in a steady state. Adding trend TFP growth is unlikely to change any of the results. Of course, it is an important assumption that climate damages do not affect TFP growth.

There are $ H=3 $ agents with identical CRRA utility functions. The coefficient of relative risk aversion, $ \sigma $, is set to $ 5 $, and the time preference factor, $ \beta $, is set to 0.74 (corresponding to an annual $\beta$ of 0.97).
Different examples of labor endowments $ \bar l_h $, $ h \in {\mathbf H} $ and market participation are considered below. 

The climate part of the model follows
~\cite{golosov2014optimal}, and it is assumed that $ \xi_2 = 0.5$ , and  $ \xi_1 = 0.4$. 
 The slow and fast depreciation rates, $ \delta_{S1} $ and $ \delta_{S2} $, are set at 1.0 and 0.97, respectively. 
Initial conditions for $ t=0 $ are  $ S_0=(118, 684) $ as the carbon in the two reservoirs
at $ t=0 $ 
and $ S=581 $ as the pre-industrial level of carbon in the atmosphere.

As explained in the introduction, there is a large disagreement in the climate science literature about the value of the equilibrium climate sensitivity (ECS). In this context, it is important to note that the ECS is not a free parameter in Earth System models but that it emerges as one characteristic of simulated climate change describing the very long-run behavior of temperature. Determining a suitable stochastic process for this parameter that can be used in macro-economic modeling is challenging, and there is no consensus on this issue. 
In this study,
the ECS-parameter, $ \lambda $, is assumed to take one of 6 possible 
values each period, 
$$ \lambda \in \{ 1.1, 2.6,  3.1, 3.6, 4.1, 5.6 \} $$ where the range from 2.6 to 4.1 is consistent the range of reasonable values  from climate science. I add extremes to 
capture additional uncertainty in the damage function. As explained in the introduction, it is  assumed that the uncertainty about $ \lambda $ reduces over time. I take all six shocks
to be equiprobable for the first two periods, $ t=0,1 $ and then set
$$ \Pe_t (z|z) = 1-0.5^{\frac{t+1}{3}} $$ and
$$ \Pe_t (z'|z) =\frac{1-\Pe_t(z|z)}{5} \mbox{ for } z' \ne z .$$
At $ t=30 $, the probability of switching climate states is therefore only about 1/1000.
There seems to be a consensus in the literature that, in the long run, the uncertainty about the ECS will be very small (see, e.g., \cite{sherwood2020assessment}).

\begin{figure}[t!]
\centering
 \includegraphics[width=12cm,height=9 cm]{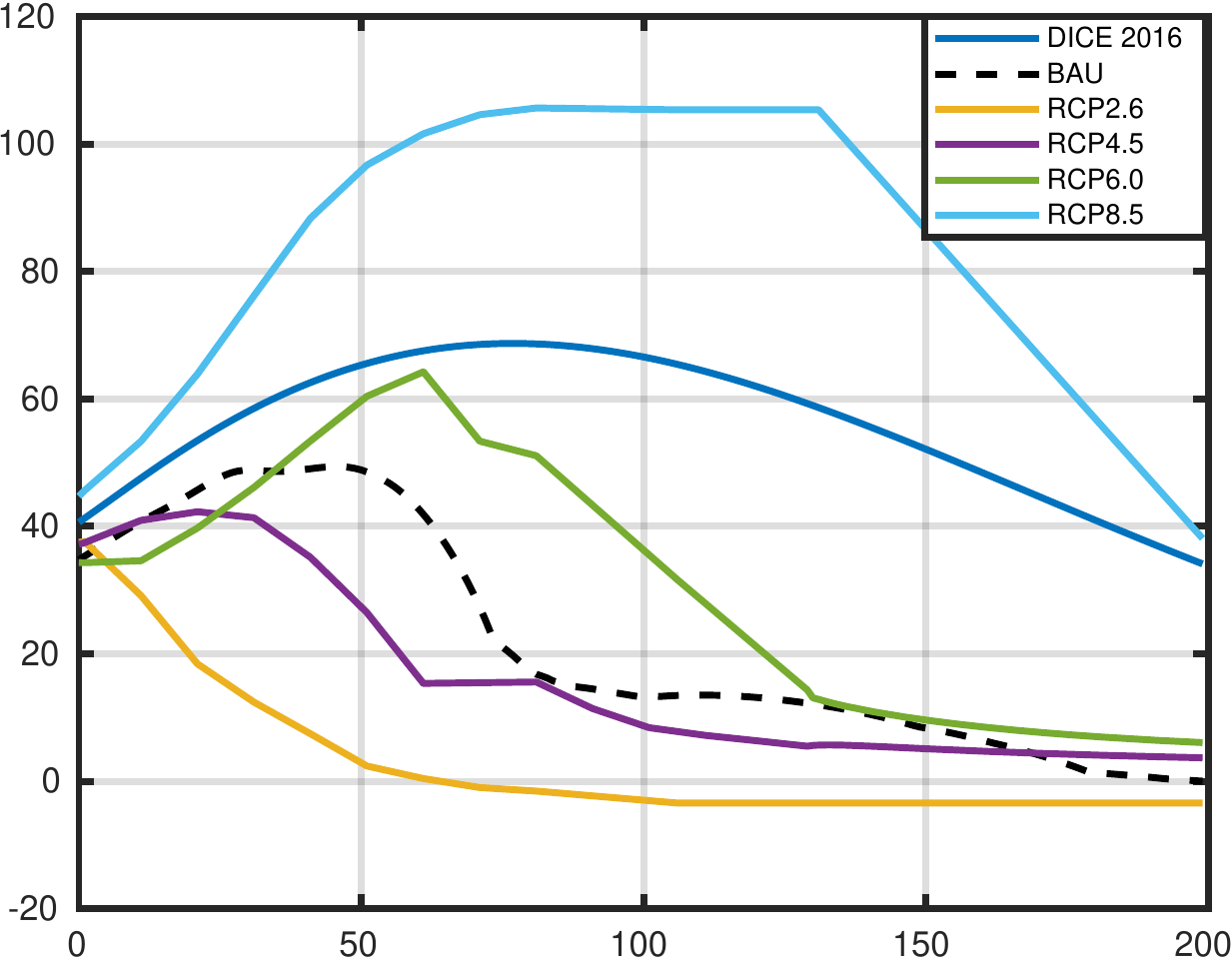}
   \caption{Global CO2 emissions (in GtCO2) in DICE 2016, the BAU-scenario in DICE 2016 and in \cite{KKPS} and the different RCP scenarios (year 0 is 2017).}
  \label{emissions_rcp-dice}
    \end{figure}
The emissions to output ratio, $ e_t $, is calibrated to match RCP 4.5 emissions over the next 300 years at steady state output.
The representative concentration pathways (RCPs) describe the results of different socioeconomic narratives that produce particular concentration profiles of greenhouse gases, aerosols, and other climatically relevant forcing agents over the 21st century, see
~\citep[][http://www.pik-potsdam.de/\textasciitilde mmalte/rcps/]{meinshausen-et-al:11}.
The RCP85 scenario, for instance, reflects a \lq\lq no policy\rq \rq \ narrative, in which total anthropogenic forcing reaches approximately $8.5 W/m^2$ in the year 2100. Conversely, the RCP2.6 scenario involves aggressive decarbonization, causing radiative forcing to peak at approximately $3 W/m^2$ around 2050 and to decline to approximately $2.6 W/m^2$ at the end of the 21st century.
 As ~\cite{hausfather2020emissions} point out, RCP8.5 was intended to explore an unlikely high-risk future, entails unrealistically high coal usage. In this light, the RCP6.0 or the RCP4.5 scenario seem more relevant for forecasting emissions absent additional abatement. The BAU scenario in \cite{nordhaus2017revisiting} leads to emissions that lie somewhere between RCP6.0 and RCP 8.5, but closer to RCP6.0 (see e.g. \cite{Folini_2021}).
In this paper, RCP4.5 emissions are assumed to describe the path of future emissions without abatement.
 The RCP4.5 scenario is a stabilization scenario, which means the radiative forcing level stabilizes before 2100 by employment of a range of technologies and strategies for reducing greenhouse gas emissions.
In the standard DICE2016 calibration there is little room for additional abatement given the parameter values assumed (see \cite{Folini_2021}). As will be shown, uncertainty together with financial frictions lead to substantial abatements on top of the ones already  present in the RCP4.5 scenario.
Figure \ref{emissions_rcp-dice} (from \cite{KKPS}) shows the emissions path for the different scenarios. Clearly, this paper starts with a very optimistic one.

The abatement technology is taken from 
~\cite{nordhaus2017revisiting} 
and it is assumed that $$ A_t(\mu_t)  = 1-\phi_{1t} \mu_t^{\phi_{2}} . $$
To calibrate the parameters $ \phi_{1t} $ and $ \phi_{2} $ one needs to take a stand on the  elasticities of substitution between dirty energy, capital, and labor, as well as on future possibilities of carbon removal.
To simplify the analysis, it is assumed in our calibration  that these coefficients are constant over time. In \cite{nordhaus2017revisiting} abatement becomes more cost-effective over time, but, as pointed out above, BAU emissions are assumed to stay relatively constant. The  RCP4.5 scenario used in this paper builds in a steep decrease of emissions starting at about 2050, and hence it feels natural to keep the abatement technology constant. It  is assumed that $ \phi_2=2 $ and $ \phi_1=0.25 $. This implies that reducing emissions by 20 percent reduces output by 1 percent, and reducing emissions by 50 percent reduces output by 10 percent. As pointed out several times, this is in addition to the large reductions already assumed in the RCP 4.5 scenario and therefore seems rather optimistic.

As~\cite{hassler2012energy} point out, assuming a low elasticity of substitution between capital, labor, and dirty energy is unrealistic when the time period is short. In fact, they find that a Leontief specification provides a good fit for annual data. However,~\cite{hassler2018consequences} make the assumption that the production function is Cobb-Douglas in capital, labor, and energy in a model, which, like the model in this paper, has periods lasting for ten years.

To calibrate damages, I use the functional form in \cite{nordhaus2017revisiting} who assumes that damages to output are given by 
$$ D(T_t) = 1-\zeta T_t^2 .$$
There is an active debate on the specification of the damage function.~\cite{hansel2020climate} strongly criticize the function posited by Nordhaus in his DICE model. They follow~\cite{howard2017few} and use Nordhaus' functional form, simply changing parameters so that damages reach $6.7$ percent of output for a 3-degree temperature increase, as opposed to only 2.1 percent in Nordhaus' calibration\footnote{It is also not clear what a good functional form for the damage function would look like, see \cite{weitzman2009modeling}. For simplicity, this paper focuses on quadratic damage functions. As \cite{Folini_2021} point out the functional form plays an important role in the determination of optimal carbon policy.}. 

As mentioned above, in this paper, climate damages are modeled as a reduction in agents' labor endowments.
I calibrate losses to the labor efficiency units to be quadratic (hence losses to output are not quadratic but instead determined by the term $ (1-\zeta T^2)^{1-\alpha} $) and take the coefficient to be $ \zeta=0.007 $. Together with the specification of the labor share, one then obtains a 4.2 percent aggregate output loss for a 3-degree temperature increase. This value lies right in the middle between Nordhaus' and more pessimistic calibration from \cite{howard2017few}.

Labor endowments and damages
 vary across agents. In the benchmark calibration, it is assumed that labor endowments are given by
$ \bar l_1=\frac{3}{4}, \ \bar l_2= \bar l_3=\frac{1}{8} $.
Agents $ h=1,2 $ trade  risky capital, while agent $h=3$ is assumed to be a hand-to-mouth consumer who consumes his labor income every period.
Agent-specific damages are
$$ D_1(T) = 1-\frac{1}{2} \zeta T^2 , \quad
 D_2(T) = 1-\zeta T^2 , \quad D_3(T)=1-4 \zeta T^2 .$$ Given the individual labor endowments across agents, the resulting individual damages always add up to aggregate damages, but, of course, damages can become very large for the third agent. This benchmark calibration is referred to as {\it BM} below.

 The fact that the hand-to-mouth agents are hardest hit by climate change plays a crucial role in the results. One could alternatively imagine a situation where, as in the main calibration, some agents suffer much more from climate change than others, but where these agents can save in risky capital.
In this alternative calibration, the roles of agents 2 and 3 are reversed, i.e., it is assumed that
$$ D_2(T) = 1-4 \zeta T^2 , \quad D_3(T)= 1-\zeta T^2. $$
Agent 2 now suffers most from climate change, but he can save in risky capital to insure against some of the damages.
This benchmark calibration is referred to as {\it AC} below.

 I also consider a third calibration where the only friction consists of market incompleteness. In that calibration, agents 1 and 2 are identical and $ \bar{1}_1=\bar{l}_2=\frac{2}{5} $ while $ \bar{l}_3=\frac{1}{5} $. Damages are
 $$ D_1(T) = D_2(T) = 1- \frac{1}{4}\zeta T^2 , \quad
 D_3(T) = 1-4 \zeta T^2 $$
The calibration is referred to as {\it IM} below.

A more careful calibration of relative damages across different individuals would require a larger number of individuals and several regions. This is beyond the scope of this paper.
\section{Results}
\label{sec:results}
As explained in the introduction, the primary focus is to explore constrained efficient taxes and transfers.
Welfare and optimal carbon taxes for an economy where all three agents can trade in a complete set of Arrow securities are compared to those in the economy with financial frictions.
Without any climate change, there would be no uncertainty, and welfare across agents would be identical for the two cases (one agent is not allowed to invest in capital, however, absent any uncertainty, the agent has no reason to do so).
However, I consider a situation where abatement is costly and where  there will always be climate change and the associated uncertainty about temperature and damages. It is then a quantitative question of how the
equilibria will differ and how optimal taxes depend on financial markets.

All results are obtained numerically. A short description of the numerical method and an error analysis can be found in the appendix.
\subsection{Constrained efficient abatement}
Table 1 reports optimal abatement levels for three different cases: A frictionless economy (CM), optimal taxes (BM), and constrained efficient taxes without transfers (NT) in the benchmark economy.
Obviously, these abatement levels vary across time and states, and they are reported for nine specific date events. It is assumed that in the first two periods, the realized shock is $ z=3 $, in the third period shock $ z= 6 $ realizes, and in the fourth period abatement is reported for all six shocks. For different histories of shocks the results are comparable, in particular in the first two periods where shocks are i.i.d. differences between shocks are somewhat small.
\begin{table}[tb!]
\begin{center}
\begin{tabular}{|c|c|c|c||c|c|c|c|c|c|}
\hline 
 \textbackslash  & $z_0=3$ & $z_1=3$ & $z_2=6$ & $z_3=1$ &
$z_3=2$ & $z_3=3$ & $ z_3=4 $ & $z_3=5 $ & $ z_3=6 $\\ 
\hline \hline
CM  & 4.49  & 5.17 & 6.69 & 4.04 & 4.43 & 4.63  & 4.88 &  5.18 & 6.96 \\ \hline
OT & 11.12 & 12.97 & 8.70 & 14.81 & 12.71 & 11.81 & 10.84 & 9.92 & 8.40 \\ \hline 
NT &  11.53 & 14.33 & 9.44 & 17.71 & 14.48 & 13.21 & 11.91 & 10.68 & 8.85 \\ \hline
\end{tabular} 
		\caption{Constrained optimal abatement}
		\label{table:1}
	\end{center}
\end{table}

Optimal abatement with financial frictions is almost always about 2.5 times higher in an economy with financial frictions than in the Pareto-optimal outcome in the frictionless economy (with the exception of $ t=2, \ z_t=6 $ and $ t=3, z_t=6 $). This seems surprising: in an optimal world, emissions,  temperature increases, and the resulting damages are significantly higher than in the sub-optimal equilibrium with financial frictions. The economic reasoning is clear; in an economy with financial friction, high carbon taxes substitute for precautionary savings, which are  impossible for the hand-to-mouth agents and only possible only in a risky asset for other agents.

In the frictionless case, optimal abatement after the fourth period (i.e. $t=3$) is higher if the economy is in a worse climate state. This makes sense since the marginal benefits of abatement are high precisely in those states. Financial frictions reverse this pattern. Optimal abatement at $ t=3 $ is highest in the low ECS states and decreases as the ECS increases. This is caused by the fact that in high ECS states at $ t=3 $, there are already substantial damages, and while the marginal benefits of abatement are large, most of the costs have to be borne by the hand-to-mouth agents who are already hard hit. If these agents could insure against this bad outcome they would be able to spend more on abatement.

A comparison between the constrained optimal case (BT) and the case without transfers (NT) shows that this is not only caused by the fact that in the optimal case, the hand-to-mouth agents have to pay a large fraction of the abatement costs. Without transfers, all agents pay proportionally to their capital holdings and their labor endowments; in this case, the hand-to-mouth agents actually only bear a small fraction of total costs. Optimal abatement is uniformly larger. In some instances, e.g. in $ z_1=3 $ or $ z_2=6 $ the difference is quite substantial and larger than 10 percent.

The issue  becomes clearer if one looks at transfers and total costs of abatement as detailed in Table \ref{table:1b}

\begin{table}[htb!]
\begin{center}
\begin{tabular}{|c|c|c|c|c|c|c|}
\hline 
 \textbackslash  & $z_3=1$ &
$z_3=2$ & $z_3=3$ & $ z_3=4 $ & $z_3=5 $ & $ z_3=6 $\\ 
\hline \hline
costs  &  1 & 0.73 & 0.63  & 0.52 &  0.43 & 0.30 \\ \hline
share  & (8, 4, 88) & (11, 5, 84) & (13,5,82) & (15,6,79) &  (17, 7, 75) & (29, 12, 59)\\ \hline 
\end{tabular} 
		\caption{Optimal abatement cost and cost-sharing}
		\label{table:1b}
	\end{center}
\end{table}
The table shows relative costs (relative to costs at $ z_3=1 $ which are normalized to one) as well as the share of these costs each agent bears in the transfer scheme. For good climate shocks with small damages, agent 3 bears almost the entire cost, and the total expenditure for abatement is high. For a bad climate shock, damages to agent 3's labor endowments are already very high (about 30 percent loss in shock six)
and the agent simply does not have the resources to finance a large abatement.

\begin{table}[htb!]
\begin{center}
\begin{tabular}{|c|c|c|c||c|c|c|c|c|c|}
\hline 
 \textbackslash  & $z_0=3$ & $z_1=3$ & $z_2=6$ & $z_3=1$ &
$z_3=2$ & $z_3=3$ & $ z_3=4 $ & $z_3=5 $ & $ z_3=6 $\\ 
\hline \hline
BM & 11.12 & 12.97 & 8.70 & 14.81 & 12.71 & 11.81 & 10.84 & 9.92 & 8.40 \\ \hline 
IM & 6.14 & 7.62 & 9.53 & 7.62 & 7.59 & 7.58 & 
7.59 & 7.64 & 9.94 \\ \hline
AC & 5.01 & 6.01 & 7.70 & 5.88  &
8.13 & 8.44 & 8.36 & 8.31 & 8.27
\\ \hline 
\end{tabular} 
		\caption{Optimal abatement, alternative calibration}
		\label{table:2}
	\end{center}
\end{table}

Of course, the magnitude of the differences between the complete markets abatement and the constrained optimal abatement with financial frictions depends crucially on the presence of hand-to-mouth consumers that are hit hardest by future damages from climate change.
While this is arguably a realistic assumption, it is useful to also examine other calibrations.
Table \ref{table:2} compares the optimal abatement in the main calibration (BM) to the optimal abatement in the alternative calibration (AC) and to optimal abatement in the calibration where the only financial friction consists of market incompleteness (IM). 
In the (AC) calibration, participation in financial markets, even if they are incomplete, leads to much lower optimal abatement levels, certainly in cases where the damages of climate change are not yet too high (in $ z_3=6 $ optimal abatement is very similar in the two cases).
Note that abatement levels are still significantly higher than in the frictionless case. 

Interestingly, in the calibration without any hand-to-mouth consumers, optimal abatement levels are again significantly higher than in (AC). This is caused by the fact that the agents that are hardest hit by climate change now constitute a significant fraction of the total population (1/5 instead of only 1/8 in the BM calibration).
The constrained optimal abatement is much larger than in the CM markets case (for this case, it is the same as in Table
\ref{table:1}, since aggregate damages remain the same). The example illustrates that even when the only friction is the incompleteness of financial markets, optimal abatement levels can differ by a lot. It should be emphasized that in this case of incomplete markets, like in the case of the complete market, abatement in the 4th period is higher in worse climate states. This is in stark contrast to the calibration with hand-to-mouth consumers, where the opposite is true. In fact, at $ z_2=6 $ and at $ z_3=6 $, constrained optimal abatement in the CM calibration is significantly higher than in the BM calibration. Even with incomplete financial markets, agents that suffer the most from climate change have enough liquidity in bad climate shocks to pay for higher abatement. 

Finally, I consider the effects of delayed implementation on optimal abatement levels.
Table \ref{table:3} shows optimal abatement for the cases where abatement is delayed for one period and or two periods (i.e., for 10 or 20 years).
Delayed implementation slightly increases optimal taxes (also, compared to the results in Table \ref{table:1}). This is also true in the frictionless case (here only reported for a delay by one period) but certainly much less so. In the low ECS states at $ t=3 $ optimal taxes in the calibration with frictions are more than three times higher than optimal taxes in the frictionless case.

\begin{table}[htb!]
\begin{center}
\begin{tabular}{|c|c|c||c|c|c|c|c|c|}
\hline 
 \textbackslash  & $z_1=3$ & $z_2=6$ & $z_3=1$ &
$z_3=2$ & $z_3=3$ & $ z_3=4 $ & $z_3=5 $ & $ z_3=6 $\\ 
\hline \hline
1, OT &  13.08 & 8.24 & 15.51 & 13.28 &  12.33 &  11.29 & 10.22 &  8.48 \\ \hline 
2, OT  & 0 & 8.99 & 16.71 & 14.19 & 13.13 & 11.97 & 10.76 & 8.60 \\ \hline 
1, CM & 5.20 & 6.72 & 4.06 & 4.46 & 4.65 & 4.90 & 5.20 & 6.98\\
\hline
\end{tabular} 
		\caption{Optimal abatement, delayed implementation}
		\label{table:3}
	\end{center}
\end{table}

To summarize the findings so far: Financial frictions have very large effects on optimal carbon taxes and optimal abatement. The less sophisticated possible transfer schemes, the higher the optimal taxes (and the lower the resulting climate damages).

\subsection{Welfare}

A natural next step is to investigate the welfare consequences of these findings. We report all welfare losses in lifetime consumption equivalent, i.e., we ask by what percentage consumption at all times and date events has to change to compensate for welfare changes caused by climate change or financial frictions.

Since the model starts off with the RCP 4.5 emissions scenario, welfare losses from climate change are generally small\footnote{In size somewhat comparable to welfare losses in DICE2016.}. Clearly, welfare losses from climate change are much larger in a very pessimistic no-policy scenario such as RCP8.5 or the BAU scenario from DICE2016. Note, however, that with discounting and moderate risk aversion welfare losses from climate change and imperfect insurance against climate change tend to be small. Potentially large welfare gains hinge critically on Inada-conditions on the utility functions.

For the three agents, I find that the
    welfare gains, in percentage, of constrained efficient carbon policy (relative to no further mitigation in RCP 4.5)  are given by 
    $$ ( 0.066, 0.486, 3.68) .$$ These are obviously smaller than total welfare losses from climate change since abatement is assumed to be costly.

    Relative to these numbers, the potential welfare gains from complete markets and optimal taxes, relative to the situation with financial frictions and optimal taxes, seem relatively large.
    With appropriate transfers that leave agent 1 at his status-quo 
    welfare level, the welfare of agents 2 and 3 can be increased by 2.82 percent by erasing all financial  frictions.
    The constrained optimal abatement in the economy with financial frictions cannot nearly compensate  for the climate risks that individuals face and cannot insure. Despite the fact that in the frictionless outcome, aggregate climate damages are significantly higher than in the benchmark, each agent's welfare is significantly higher because the individual risks can be diversified away.

    If abatement is impossible, the welfare losses from financial frictions are naturally significantly larger. Aggregate damages now are very similar, and the larger individual risks in the economy with frictions have relatively large effects on welfare.
    Welfare gains from completing the markets, together with transfers that keep agent 1 at his status-quo welfare and ensure uniform gains for agents 2 and 3, lead to welfare gains for these agents of 
     4.01 percent.

Lastly, welfare for the optimal abatement case is compared to that in the case with efficient taxes but without any transfers. Moving to such a scheme implies welfare changes (in percent) for the three agents of 
$$ (-0.331,-0.290, 2.149) .$$ Not surprisingly, Agent 3, who are most affected by climate damages and has no access to financial markets, gains a lot if he no longer has to pay a large fraction of the abatement. However, agent 2 and agent 1 incur significant losses.
As before, since abatement is higher in the scenario than in the {\it constrained optimal} policy (and much higher than in the optimal policy with complete markets), climate change damages are lowest in this case. Nevertheless, welfare losses with respect to the complete markets optimum are substantial, in fact, slightly (about 0.01 percent) higher than in the constrained optimal case. 
\section{Conclusion}
\label{sec:concl}
Climate change poses large aggregate risks that will likely have very different effects on different households. Unfortunately, these risks cannot be shared in existing financial markets. I give some quantitative examples to show that the inability to share these risks leads to much higher abatement and lower emissions at a constrained optimum. 

In the calibrated examples, I focus on uncertainty about future temperature and aggregate damages.
Of course, there might be many other risks caused by anthropogenic climate change which cannot be shared in markets and have significant effects on agents' well beings. Risks associated with tipping points have played a prominent role in the literature (e.g., \cite{lemoine2014watch}, \cite{cai2013social}), but it might  be as important to model the effects of climate change on idiosyncratic income risk. It is subject to further research to understand how these uncertainties interact and how large the quantitative impact on welfare and optimal taxes are.

This paper focuses on heterogeneity across consumers. Another important factor in optimal carbon taxation lies in heterogeneity across producers, particularly across regions.\\

\bibliography{bib_econ.bib}{}

\begin{thebibliography}{47}
\providecommand{\natexlab}[1]{#1}
\providecommand{\url}[1]{\texttt{#1}}
\expandafter\ifx\csname urlstyle\endcsname\relax
  \providecommand{\doi}[1]{doi: #1}\else
  \providecommand{\doi}{doi: \begingroup \urlstyle{rm}\Url}\fi

\bibitem[Barnett et~al.(2020)Barnett, Brock, and Hansen]{barnett2020pricing}
Michael Barnett, William Brock, and Lars~Peter Hansen.
\newblock Pricing uncertainty induced by climate change.
\newblock \emph{The Review of Financial Studies}, 33\penalty0 (3):\penalty0
  1024--1066, 2020.

\bibitem[Barnett et~al.(2022{\natexlab{a}})Barnett, Brock, and
  Hansen]{barnett2022aclimate}
Michael Barnett, William Brock, and Lars~Peter Hansen.
\newblock Confronting uncertainty in the climate change dynamics.
\newblock \emph{working paper}, 2022{\natexlab{a}}.

\bibitem[Barnett et~al.(2022{\natexlab{b}})Barnett, Brock, and
  Hansen]{barnett2022climate}
Michael Barnett, William Brock, and Lars~Peter Hansen.
\newblock Climate change uncertainty spillover in the macroeconomy.
\newblock \emph{NBER Macroeconomics Annual}, 36\penalty0 (1):\penalty0
  253--320, 2022{\natexlab{b}}.

\bibitem[Barrage(2020)]{barrage2020optimal}
Lint Barrage.
\newblock Optimal dynamic carbon taxes in a climate--economy model with
  distortionary fiscal policy.
\newblock \emph{The Review of Economic Studies}, 87\penalty0 (1):\penalty0
  1--39, 2020.

\bibitem[Brumm et~al.(2017)Brumm, Kryczka, and Kubler]{brumm2017recursive}
Johannes Brumm, Dominika Kryczka, and Felix Kubler.
\newblock Recursive equilibria in dynamic economies with stochastic production.
\newblock \emph{Econometrica}, 85\penalty0 (5):\penalty0 1467--1499, 2017.

\bibitem[Cai(2020)]{cai2020role}
Yongyang Cai.
\newblock The role of uncertainty in controlling climate change.
\newblock \emph{arXiv preprint arXiv:2003.01615}, 2020.

\bibitem[Cai et~al.(2013)Cai, Judd, and Lontzek]{cai2013social}
Yongyang Cai, Kenneth~L Judd, and Thomas~S Lontzek.
\newblock The social cost of stochastic and irreversible climate change.
\newblock Technical report, National Bureau of Economic Research, 2013.

\bibitem[Carattini et~al.(2021)Carattini, Heutel, and
  Melkadze]{carattini2021climate}
Stefano Carattini, Garth Heutel, and Givi Melkadze.
\newblock Climate policy, financial frictions, and transition risk.
\newblock Technical report, National Bureau of Economic Research, 2021.

\bibitem[Chichilnisky and Heal(1994)]{chichilnisky1994should}
Graciela Chichilnisky and Geoffrey Heal.
\newblock Who should abate carbon emissions?: An international viewpoint.
\newblock \emph{Economics Letters}, 44\penalty0 (4):\penalty0 443--449, 1994.

\bibitem[Cruz and Rossi-Hansberg(2021)]{Cruz2021}
Jose-Luis Cruz and Esteban Rossi-Hansberg.
\newblock {The Economic Geography of Global Warming}.
\newblock 2021.

\bibitem[Davila et~al.(2012)Davila, Hong, Krusell, and
  R{\'\i}os-Rull]{davila2012constrained}
Julio Davila, Jay~H Hong, Per Krusell, and Jos{\'e}-V{\'\i}ctor R{\'\i}os-Rull.
\newblock Constrained efficiency in the neoclassical growth model with
  uninsurable idiosyncratic shocks.
\newblock \emph{Econometrica}, 80\penalty0 (6):\penalty0 2431--2467, 2012.

\bibitem[Dennig et~al.(2015)Dennig, Budolfson, Fleurbaey, Siebert, and
  Socolow]{dennig2015inequality}
Francis Dennig, Mark~B Budolfson, Marc Fleurbaey, Asher Siebert, and Robert~H
  Socolow.
\newblock Inequality, climate impacts on the future poor, and carbon prices.
\newblock \emph{Proceedings of the National Academy of Sciences}, 112\penalty0
  (52):\penalty0 15827--15832, 2015.

\bibitem[Dierker et~al.(2002)Dierker, Dierker, and
  Grodal]{dierker2002nonexistence}
Egbert Dierker, Hildegard Dierker, and Birgit Grodal.
\newblock Nonexistence of constrained efficient equilibria when markets are
  incomplete.
\newblock \emph{Econometrica}, 70\penalty0 (3):\penalty0 1245--1251, 2002.

\bibitem[Douenne et~al.(2022)Douenne, Hummel, and Pedroni]{douenne2022optimal}
Thomas Douenne, Albert~Jan Hummel, and Marcelo Pedroni.
\newblock Optimal fiscal policy in a climate-economy model with heterogeneous
  households.
\newblock 2022.

\bibitem[Foley(1970)]{foley1970lindahl}
Duncan~K Foley.
\newblock Lindahl's solution and the core of an economy with public goods.
\newblock \emph{Econometrica: Journal of the Econometric Society}, pages
  66--72, 1970.

\bibitem[Folini et~al.(2021)Folini, Kubler, Malova, and
  Scheidegger]{Folini_2021}
Doris Folini, Felix Kubler, Aleksandra Malova, and Simon Scheidegger.
\newblock The climate in climate economics.
\newblock \emph{Available at SSRN 3885021}, 2021.

\bibitem[Gillingham et~al.(2015)Gillingham, Nordhaus, Anthoff, Blanford,
  Bosetti, Christensen, McJeon, Reilly, and Sztorc]{gillingham2015modeling}
Kenneth Gillingham, William~D Nordhaus, David Anthoff, Geoffrey Blanford,
  Valentina Bosetti, Peter Christensen, Haewon McJeon, John Reilly, and Paul
  Sztorc.
\newblock Modeling uncertainty in climate change: A multi-model comparison.
\newblock Technical report, National Bureau of Economic Research, 2015.

\bibitem[Gollier(2001)]{gollier2001wealth}
Christian Gollier.
\newblock Wealth inequality and asset pricing.
\newblock \emph{The Review of Economic Studies}, 68\penalty0 (1):\penalty0
  181--203, 2001.

\bibitem[Golosov et~al.(2014)Golosov, Hassler, Krusell, and
  Tsyvinski]{golosov2014optimal}
Mikhail Golosov, John Hassler, Per Krusell, and Aleh Tsyvinski.
\newblock Optimal taxes on fossil fuel in general equilibrium.
\newblock \emph{Econometrica}, 82\penalty0 (1):\penalty0 41--88, 2014.

\bibitem[H{\"a}nsel et~al.(2020)H{\"a}nsel, Drupp, Johansson, Nesje, Azar,
  Freeman, Groom, and Sterner]{hansel2020climate}
Martin~C H{\"a}nsel, Moritz~A Drupp, Daniel~JA Johansson, Frikk Nesje,
  Christian Azar, Mark~C Freeman, Ben Groom, and Thomas Sterner.
\newblock Climate economics support for the un climate targets.
\newblock \emph{Nature Climate Change}, pages 1--9, 2020.

\bibitem[Hassler et~al.(2012)Hassler, Krusell, and Olovsson]{hassler2012energy}
John Hassler, Per Krusell, and Conny Olovsson.
\newblock Energy-saving technical change.
\newblock Technical report, National Bureau of Economic Research, 2012.

\bibitem[Hassler et~al.(2018)Hassler, Krusell, and
  Olovsson]{hassler2018consequences}
John Hassler, Per Krusell, and Conny Olovsson.
\newblock The consequences of uncertainty: climate sensitivity and economic
  sensitivity to the climate.
\newblock \emph{Annual Review of Economics}, 10:\penalty0 189--205, 2018.

\bibitem[Hausfather and Peters(2020)]{hausfather2020emissions}
Zeke Hausfather and Glen~P Peters.
\newblock Emissions--the ‘business as usual’story is misleading, 2020.

\bibitem[Howard and Sterner(2017)]{howard2017few}
Peter~H Howard and Thomas Sterner.
\newblock Few and not so far between: a meta-analysis of climate damage
  estimates.
\newblock \emph{Environmental and Resource Economics}, 68\penalty0
  (1):\penalty0 197--225, 2017.

\bibitem[Jensen and Traeger(2014)]{jensen2014optimal}
Svenn Jensen and Christian~P Traeger.
\newblock Optimal climate change mitigation under long-term growth uncertainty:
  Stochastic integrated assessment and analytic findings.
\newblock \emph{European Economic Review}, 69:\penalty0 104--125, 2014.

\bibitem[Kaneko(1977)]{kaneko1977ratio}
Mamoru Kaneko.
\newblock The ratio equilibrium and a voting game in a public goods economy.
\newblock \emph{Journal of Economic Theory}, 16\penalty0 (2):\penalty0
  123--136, 1977.

\bibitem[Klein et~al.(2008)Klein, Krusell, and Rios-Rull]{klein2008time}
Paul Klein, Per Krusell, and Jose-Victor Rios-Rull.
\newblock Time-consistent public policy.
\newblock \emph{The Review of Economic Studies}, 75\penalty0 (3):\penalty0
  789--808, 2008.

\bibitem[Knutti et~al.(2017)Knutti, Rugenstein, and Hegerl]{knutti2017beyond}
Reto Knutti, Maria~AA Rugenstein, and Gabriele~C Hegerl.
\newblock Beyond equilibrium climate sensitivity.
\newblock \emph{Nature Geoscience}, 10\penalty0 (10):\penalty0 727--736, 2017.

\bibitem[Kotlikoff et~al.(2021{\natexlab{a}})Kotlikoff, Kubler, Polbin, and
  Scheidegger]{kotlikoff2020pareto}
Laurence Kotlikoff, Felix Kubler, Andrey Polbin, and Simon Scheidegger.
\newblock {Pareto-Improving Carbon-Risk Taxation}.
\newblock \emph{Economic Policy}, 02 2021{\natexlab{a}}.
\newblock ISSN 0266-4658.
\newblock \doi{10.1093/epolic/eiab008}.
\newblock URL \url{https://doi.org/10.1093/epolic/eiab008}.
\newblock eiab008.

\bibitem[Kotlikoff et~al.(2021{\natexlab{b}})Kotlikoff, Kubler, Polbin, and
  Scheidegger]{KKPS}
Laurence~J Kotlikoff, Felix Kubler, Andrey Polbin, and Simon Scheidegger.
\newblock Can today's and tomorrow's world uniformly gain from carbon taxation?
\newblock Working Paper 29224, National Bureau of Economic Research, September
  2021{\natexlab{b}}.
\newblock URL \url{http://www.nber.org/papers/w29224}.

\bibitem[Krusell and Smith~Jr(2022)]{krusell2022climate}
Per Krusell and Anthony~A Smith~Jr.
\newblock Climate change around the world.
\newblock Technical report, National Bureau of Economic Research, 2022.

\bibitem[Lemoine and Traeger(2014)]{lemoine2014watch}
Derek Lemoine and Christian Traeger.
\newblock Watch your step: optimal policy in a tipping climate.
\newblock \emph{American Economic Journal: Economic Policy}, 6\penalty0
  (1):\penalty0 137--66, 2014.

\bibitem[Malin et~al.(2011)Malin, Krueger, and Kubler]{malin2011solving}
Benjamin~A Malin, Dirk Krueger, and Felix Kubler.
\newblock Solving the multi-country real business cycle model using a
  smolyak-collocation method.
\newblock \emph{Journal of Economic Dynamics and Control}, 35\penalty0
  (2):\penalty0 229--239, 2011.

\bibitem[Mas-Colell and Silvestre(1989)]{mas1989cost}
Andreu Mas-Colell and Joaquim Silvestre.
\newblock Cost share equilibria: A lindahlian approach.
\newblock \emph{Journal of Economic Theory}, 47\penalty0 (2):\penalty0
  239--256, 1989.

\bibitem[{Meinshausen} et~al.(2011){Meinshausen}, {Raper}, and
  {Wigley}]{meinshausen-et-al:11}
M.~{Meinshausen}, S.~C.~B. {Raper}, and T.~M.~L. {Wigley}.
\newblock {Emulating coupled atmosphere-ocean and carbon cycle models with a
  simpler model, MAGICC6 - Part 1: Model description and calibration}.
\newblock \emph{Atmospheric Chemistry \& Physics}, 11:\penalty0 1417--1456,
  February 2011.
\newblock \doi{10.5194/acp-11-1417-2011}.

\bibitem[Mor{\'e} and Wild(2014)]{more2014you}
Jorge~J Mor{\'e} and Stefan~M Wild.
\newblock Do you trust derivatives or differences?
\newblock \emph{Journal of Computational Physics}, 273:\penalty0 268--277,
  2014.

\bibitem[Moulin(1987)]{moulin1987egalitarian}
Herve Moulin.
\newblock Egalitarian-equivalent cost sharing of a public good.
\newblock \emph{Econometrica: Journal of the Econometric Society}, pages
  963--976, 1987.

\bibitem[Nordhaus(2017)]{nordhaus2017revisiting}
William~D Nordhaus.
\newblock Revisiting the social cost of carbon.
\newblock \emph{Proceedings of the National Academy of Sciences}, page
  201609244, 2017.

\bibitem[Nordhaus and Yang(1996)]{nordhaus1996regional}
William~D Nordhaus and Zili Yang.
\newblock A regional dynamic general-equilibrium model of alternative
  climate-change strategies.
\newblock \emph{The American Economic Review}, pages 741--765, 1996.

\bibitem[Pigou(1920)]{pigou1920wealth}
Arthur~C Pigou.
\newblock Wealth and welfare, london, 1912.
\newblock \emph{Later editions, The Eco$\neg$ nomics of Welfare}, 1920.

\bibitem[Roe and Baker(2007)]{roe2007climate}
Gerard~H Roe and Marcia~B Baker.
\newblock Why is climate sensitivity so unpredictable?
\newblock \emph{Science}, 318\penalty0 (5850):\penalty0 629--632, 2007.

\bibitem[Sherwood et~al.(2020)Sherwood, Webb, Annan, Armour, Forster,
  Hargreaves, Hegerl, Klein, Marvel, Rohling, et~al.]{sherwood2020assessment}
SC~Sherwood, Mark~J Webb, James~D Annan, Kyle~C Armour, Piers~M Forster,
  Julia~C Hargreaves, Gabriele Hegerl, Stephen~A Klein, Kate~D Marvel, Eelco~J
  Rohling, et~al.
\newblock An assessment of earth's climate sensitivity using multiple lines of
  evidence.
\newblock \emph{Reviews of Geophysics}, 58\penalty0 (4):\penalty0
  e2019RG000678, 2020.

\bibitem[Starr(1973)]{starr1973optimal}
Ross~M Starr.
\newblock Optimal production and allocation under uncertainty.
\newblock \emph{The Quarterly Journal of Economics}, 87\penalty0 (1):\penalty0
  81--95, 1973.

\bibitem[Traeger(2019)]{traeger2018ace}
Christian~P Traeger.
\newblock Ace--analytic climate economy (with temperature and uncertainty).
\newblock \emph{Working paper}, 2019.

\bibitem[Van~den Bremer and Van~der Ploeg(2021)]{van2021risk}
Ton~S Van~den Bremer and Frederick Van~der Ploeg.
\newblock The risk-adjusted carbon price.
\newblock \emph{American Economic Review}, 111\penalty0 (9):\penalty0
  2782--2810, 2021.

\bibitem[Weitzman(2009)]{weitzman2009modeling}
Martin~L Weitzman.
\newblock On modeling and interpreting the economics of catastrophic climate
  change.
\newblock \emph{The Review of Economics and Statistics}, 91\penalty0
  (1):\penalty0 1--19, 2009.

\bibitem[Zaliapin and Ghil(2011)]{Zaliapin.Ghil.2011}
Ilya Zaliapin and Michael Ghil.
\newblock Reply to g.h. roe{\textquoteright}s and m.b. baker{\textquoteright}s
  comment on "another look at climate sensitivity".
\newblock \emph{Nonlinear Processes in Geophysics}, 18\penalty0 (1):\penalty0
  129{\textendash}131, feb 2011.
\newblock \doi{10.5194/npg-18-129-2011}.

\end{thebibliography}
\bibliographystyle{apalike}
\begin{appendices}
\section{Computational details}
It is assumed that recursive equilibrium with constrained optimal carbon policy can be characterized by a system of functional equations involving unknown policy functions but also value functions.
To solve this system numerically, I use standard projection methods (see, e.g., \cite{malin2011solving}). The endogenous state space consists of the capital holdings of agents $ h=1,2 $ and of the two CO2 reservoirs $ S_1, S_2 $ and is, therefore, four-dimensional. Since there are no occasionally binding constraints, Smolyak-polynomials, as in \cite{malin2011solving}, are ideally suited for approximating the unknown functions.
The maximal errors in agents' Euler equations are below $ 10^{-3} $ in all computations. Unfortunately, another potential source of numerical error is the numerical computation of the optimal tax.
For this, one needs to estimate the derivative of the agents' value function with respect to CO2 concentrations in the atmosphere.
The error in the computation of the level of agents' value functions can be controlled by verifying the quality of the numerical approximation in each iteration. The derivative needs to be computed using finite differences. The method from
\cite{more2014you} is used to estimate the noise in the value function and to determine the optimal step length for the finite-difference approximation. Obtaining an estimate for the error in the level of optimal abatement appears difficult.

\end{appendices}
\end{document}